\documentclass{kluwer}

\newcommand{\be}{\begin{equation}}     
\newcommand{\ee}{\end{equation}}
\newcommand{\ba}{\begin{eqnarray}}
\newcommand{\ea}{\end{eqnarray}}

\catcode`@=11
\newcommand\eqalign[1]{\null\,\vcenter{\openup\jot\m@th
    \ialign{\strut\hfil$\displaystyle{##}$&$\displaystyle{{}##}$\hfil
        \crcr#1\crcr}}\,}

\begin{document}
\begin{article}
\begin{opening}
\title{Stability of axial orbits in galactic potentials}
\author{Cinzia \surname{Belmonte}$^{(1)}$}
\author{Dino \surname{Boccaletti}$^{(2)}$}
\author{Giuseppe \surname{Pucacco}$^{(3)}$}
\runningauthor{C. Belmonte, D. Boccaletti and G. Pucacco}
\runningtitle{Stability of axial orbits in galactic potentials}
\institute{$^{(1)}$ Dipartimento di Fisica -- Universit\`a di Roma ``la Sapienza"\\
$^{(2)}$ Dipartimento di Matematica -- Universit\`a di Roma ``la Sapienza"\\
$^{(3)}$ Dipartimento di Fisica -- Universit\`a di Roma ``Tor Vergata"
and INFN -- Sezione Tor Vergata}

\begin{abstract}
We investigate the dynamics in a galactic potential with two reflection symmetries. The phase-space structure of the real system is approximated with a resonant detuned normal form constructed with the method based on the Lie transform. Attention is focused on the stability properties of the axial periodic orbits that play an important role in galactic models. Using energy and ellipticity as parameters, we find analytical expressions of bifurcations and compare them with numerical results available in the literature.

\end{abstract}

\end{opening}


\section{Introduction}
To determine salient features of the orbital structure of non-integrable potentials is an important topic in dynamical astronomy. Techniques based on the various versions of perturbation theory have been applied to several examples and with various degrees of approximation (for a review, see, e.g., Contopoulos, 2002). Of particular interest is to understand motion in potentials which seem to be suitable to describe elliptical galaxies. Among other features, the knowledge of the stability properties of the main periodic orbits is of paramount importance, since the bulk of density distribution is shaped by the stars in regular phase-space regions around stable periodic orbits (Binney and Tremaine, 1987). In particular, for triaxial ellipsoids, periodic orbits along symmetry axes play a special role. 
An enormous effort has therefore been devoted to investigate families and bifurcations of periodic orbits, starting with the study of models based on perturbed oscillators (again, for a review, Contopoulos, 2002) and gradually exploring more realistic galactic potentials with numerical (Miralda-Escud\'e and Schwarzschild, 1989; Fridman and Merritt, 1997) and semi--analytical (de Zeeuw and Merritt, 1983; Scuflaire, 1995) approaches. 

One of the most powerful analytic tools is the normal form approximation of a non integrable system. Although the normal form approach is quite widespread in galactic dynamics, its use in studying stability of periodic orbits has not been as systematic as the theory could allow (Sanders and Verhulst, 1985). Aim of the present paper is to apply the Lie transform normalization method (Dragt and Finn, 1976; Finn, 1984; Koseleff, 1994) to approximate the dynamics of a Binney logarithmic potential (Binney and Tremaine, 1987). We compare the findings to that of Miralda-Escud\'e and Schwarzschild (1989), who employ purely numerical techniques to implement the Floquet method and to that of Scuflaire (1995), who studies the stability of axial orbits by solving the Hill-like perturbation equation with the Lindstedt--Poincar\'e approach. Another example that is briefly treated is provided by the galactic Schwarzschild (1979) potential with a comparison to the results of de Zeeuw and Merritt (1983). These authors based their approach on the averaging procedure of normalization: it is therefore interesting a comparison with that method also. We remark that, in careful numerical computations, the accuracy of predictions is usually much higher than in approximate analytical approaches. However, a reliable analytic tool is of invaluable help to gather a global overview of the behavior of the system. 

To study the linear orbital stability of the main periodic orbits with a truncated normal form one can proceed in essentially two ways: the most general and exhaustive is that of determining the explicit form of the normal modes and solve the equation of their perturbation. A less general but easier approach is that of determining the nature of the fixed points on a surface of section. This is constructed with the aid of the approximate integral of motion provided by the normalization. 
The first method is in general quite cumbersome and can be applied when the procedure of reduction to a single degree of freedom Hamiltonian system and the use of action--angle variables lead to a reasonably simple system of equations. The second one is clearly less general but relies on simple geometric arguments related to the Hessian of a polynomial in its critical points and is, at least in principle, quite easy to implement. In this work we are going to apply both methods to perform the comparison mentioned in the paragraph above.

In galactic dynamics, the periodic orbits along the axes of symmetry (axial orbits) play a particularly important role; moreover they are easily identified both as normal modes of the reduced system and as ``central'' fixed points on the surfaces of section. Therefore, we will limit the detailed evaluation of the stability characteristics in the parameter space to these axial orbits. However, both procedures we have followed are quite general and can be directly applied to all periodic orbits of sufficiently low commensurability. 

From the results obtained, we can state that the predictive power of the normal form ranges well outside the neighborhood in which the expansion of the original Hamiltonian is performed. It is rather related to the extent of the asymptotic convergence radius of the approximate integrals of motion. However, in concrete applications, the validity of the prediction has to be corroborated with an independent evaluation of the best suited resonant normal form of the problem at hand. A criterion for the choice is illustrated in the last section devoted to the applications and is connected to the ratio between the frequency of the periodic orbit and that of a normal perturbation to it.

The plan of the paper is as follows: in Section 2 we recall the procedure of normalization as applied to reflection symmetry potentials; in Sections 3 and 4 we study the 1:1 and 1:2 resonances respectively; in Section 5 we reconstruct approximate integrals of motion; in Section 6 we compare our analytical results with those available in the literature.

\section{The procedure}

\subsection{General}

Suppose the original system is given by a Hamiltonian
\be
   H({\bf p,q})= \frac{1}{2}(p_x^2+p_y^2) + V(x^2,y^2),
 \ee
with $V$ a smooth potential with an absolute minimum and reflection symmetry with respect to both axes. In our case we will use the Schwarzschild and logarithmic potentials described below. We expand the potential up to some given degree so that 
\be\label{potexp}
   V( x,y;\varepsilon)= 
   \sum_{n=0}^{\infty} \varepsilon^n V_n( x,y)
\ee
and look for a new Hamiltonian given by 
\begin{equation}\label{HK}
     K({\bf P},{\bf Q};\varepsilon)=\sum_{n=0}^{\infty}\varepsilon^n K_n ({\bf P},{\bf Q}; \varepsilon)=M_g^{-1}H({\bf p},{\bf q};\varepsilon)\, ,
  \end{equation}
  where ${\bf P},{\bf Q}$ result from the canonical transformation
  \be
  ({\bf P},{\bf Q}) = M_g ({\bf p},{\bf q}).\ee
By expanding (\ref{HK}) in power series of $\varepsilon$ and equating the coefficients of the same order, one has
\be\label{EHK} \begin{array}{ll}
    K_0=&H_0 \, ,\\ \\
   &\vdots \\ \\
    K_n=&H_n +\sum_{m=0}^{n-1}M_{n-m}H_m = H_n +M_nH_0 +\sum_{m=1}^{n-1}M_{n-m}H_m \, ,\\ \\
   &\vdots 
\end{array}\ee
The linear differential operator $M_g$ is defined by
\begin{equation}\label{eqn:OperD-F}
    M_g\equiv e^{-\varepsilon L_{g_1}}e^{-\varepsilon^2 L_{g_2}}\cdots e^{-\varepsilon^n L_{g_n}}\cdots,
\end{equation}
where the functions $g_n$ are the coefficients in the expansion of the generating function of the canonical transformation and the linear differential operator $L_{S}$ is defined through the Poisson bracket
\begin{equation} \label{eq:OdL}
    L_S f \equiv \{S,f\} \equiv \sum_{l=1}^{2}\left(
    {{\partial{S}} \over {\partial {q_l}}}
    {{\partial{f}} \over {\partial {p_l}}} - 
    {{\partial{S}} \over {\partial {p_l}}}
    {{\partial{f}} \over {\partial {q_l}}}\right).
\end{equation}
The exponentials in the definition of $M_g$ are intended as the formal sum of a power series so that it gives rise to a near identity coordinate transformation known as Lie series.

The unperturbed part of the Hamiltonian, $H_{0}$, determines the form of the transformation. In fact, the new Hamiltonian $K$ is said to be {\it in normal form} if
\be
\{H_0,K\}=0.
\ee
This condition is used at each step of the procedure to determine each function $g_n$ in order to eliminate as much as possible terms in the new Hamiltonian. The only terms of which $K$ is made of are those staying in the kernel of the operator $L_{H_{0}}$ associated to $H_{0}$ through the definition above. The procedure is stopped at some ``optimal'' order and therefore in all ensuing discussion we refer to a ``truncated'' normal form.
$H_0$ must be considered a function of the new coordinates at each step in the process: it is therefore an integral of the motion for the new Hamiltonian $K$. The function 
\be
I = K - H_{0} \ee
can be therefore used as a second integral of the motion conveying approximate informations on the dynamics of the original system. For practical applications (for example to compare results with numerical computations) it is useful to express approximating functions in the original physical coordinates. Inverting the coordinate transformation, the new integral of motion can be expressed in terms of the original variables. Denoting it as the power series
\be I = \sum_{n=0}^{\infty}\varepsilon^n I_n,\ee
its terms can be recovered by means of 
\be
I_n=H_n-K_n +\sum_{m=1}^{n-1}M_{n-m}\big[H_m-I_m\big] \, , \qquad n\geq 1\, . 
\label{eqn:In}\ee
We remark that in all subsequent applications involving series expansions, the role of the perturbation parameter can also be played by the size of the neighbourhood of the origin where the Hamiltonian is considered. Therefore the powers of the parameter $\varepsilon$ are left in all expansion formulas just to indicate their order and are treated as unity in the computations.

\subsection{Galactic potentials}

The model potentials we will consider are the Binney logarithmic potential (Binney and Tremaine, 1987) and the Schwarzschild (1979) potential. In both cases we will actually need an expansion of the form
(\ref{potexp}) and we will assume that each term can be written as a homogeneous polynomial of degree $k+2$ of the form
\be
V_k( x,y) = \sum_{j=0}^{k+2} \frac{1}{k+2} a_{(j,k+2-j)} x^{j} y^{k+2-j}.\ee
The logarithmic potential is 
\be\label{LP}
 V = \frac12 \log(1 + x^2 + y^2 / q^2)\ee
 and plays a very important role in galactic dynamics because, despite its very simple form, it has realistic features like a density distribution compatible at large radii with flat rotation curves. The form written here is simplified by the choice of fixing the length scale (the ``core radius'' $R_{c}$) equal to one, but this is not a limitation due to the invariance in both the length scale and the energy scale. With these units, the energy $E$ may take any non negative value 
 \be 0 \le E < \infty. \ee
 The parameter giving the ``ellipticity'' of the figure ranges in the interval
 \be
 0.6 \le q \le 1. \ee
 Lower values of $q$ can in principle be considered but correspond to a non physical density distribution. Values greater than unity are included in the treatment by reversing the role of the coordinate axes.
 The series expansion of the logarithmic potential is 
\be
 V = \sum_{j=1}^{\infty} \frac{(-1)^{j+1}}{2j} (x^2 + y^2 / q^2)^{j},
 \ee
so that the lowest order coefficients are
\ba
a_{20} &=& \omega_1^{2}=1,\quad a_{02} = \omega_2^{2}=1/q^{2}, \label{alog1} \\
a_{40} &=& -1,\quad
a_{22} = -1/q^2,\quad
a_{04} = -1/q^4, \label{alog2} \\
a_{60} &=& 1,\quad
a_{42} = 3/q^2,\quad
a_{24} = 3/q^4,\quad
a_{06} = 1/q^{6}. \label{alog3} \ea
The Schwarzschild (1979) potential is to be considered more for its historical role rather than for its practical usefulness. However, it has been deeply investigated and is therefore a good benchmark for comparison. It can be written as
\be
V = u(r) + \frac{x^{2}-y^{2}}{r^{2}} w(r) + 1,\ee
where 
\ba
u &=& - \frac1r \log \left(\sqrt{1+r^{2}}+r\right) - 
    c_{1} \frac{r^{2}}{2(1 + c_{2} r^{2})^{3/2}},\\
w &=& -c_{3} \frac{r^{2}}{(1 + c_{4} r^{2})^{3/2}},\ea
are two functions of $r = \sqrt{x^{2}+y^{2}}$ such that $0 < u,w < 1$ and the $c$'s are fixed constants. With the choice of de Zeeuw and Merritt (1983)\footnote{Note the correction in $c_{4}$ with respect to the value reported in the appendix of de Zeeuw and Merritt (1983), necessary to comply with the other reported constants.}
\be
c_{1} = 0.064,\quad
c_{2} = 0.655,\quad
c_{3} = 0.015,\quad
c_{4} = 0.481,\ee
the lowest order coefficients are
\ba
\omega_1 &=&0.421,\quad
\omega_2 =0.601, \label{SA1}\\
a_{40} &=& -0.042,\quad
a_{22} = -0.174, \quad
a_{04} = -0.307, \label{SA2} \\
a_{60} &=& -0.006, \quad
a_{42} = 0.221, \quad
a_{24} = 0.460, \quad
a_{06} = 0.233.\ea
The energy range in the Schwarzschild potential is
\be
 0 \le E \le 1. \ee
 
\subsection{Detuning the normal form}

The natural setting in which one can perform a low order normalization is therefore that of a perturbed quadratic Hamiltonian with a potential starting  with
a harmonic term. In the general case in which the frequencies are rationally independent, the kernel of the operator associated to
\be
H_{0} = \frac12 (p_x^2 + p_y^2 + \omega_1^2 x^2 + \omega_2^2 y^2)
\ee
is trivial, consisting only of functions of the partial energies in the harmonic potential: it is customary to refer to the normal form constructed in this case as a ``Birkhoff'' normal form (Birkhoff, 1927). The presence of terms with small denominators in the expansion, forbids in general its convergence. It is therefore more effective to work since the start with a ``resonant'' normal form, with  a ``richer'' kernel that allows to reconstruct the main natural resonances shaping the phase-space of the system. To catch the main features of the orbital structure, we therefore approximate the frequencies with an integer ratio plus a small ``detuning'' that we assume ${\rm O}(\varepsilon^2)$
\be\label{DET}
\frac{\omega_1}{\omega_2} = \frac{m}{n} + \varepsilon^2 \delta\ee
and we speak of a {\it detuned} ({\it m:n}) {\it resonance}, with $m+n$ the {\it order} of the resonance.

We have to put the system in a form suitable to apply the normalization procedure: we rescale variables in order to put the Hamiltonian in the form
\be\label{detH}
H = \frac12 [(m+ n \delta) (p_x^2+x^2) + n (p_y^2+y^2)]+ \sum_{k=2}^{\infty}\sum_{j=0}^{k+2} \frac{b_{(j,k+2-j)}}{k+2} x^{j} y^{k+2-j}\ee
where we have used the same notation for the rescaled variables and
\be \label{abc}
b_{(j,k+2-j)}=  \frac{n a_{(j,k+2-j)}}{\omega_1^{j/2} \omega_2^{(k+4-j)/2}} . \ee
The procedure is now that of an ordinary resonant ``Birkhoff--Gustavson'' normalization (Gustavson, 1966; Moser, 1968) with two variants: the coordinate transformations are performed through the Lie series and the detuning quadratic term is treated as a term of higher order and put in the perturbation.

\subsection{Choice of the resonance}

Given an arbitrary pair of unperturbed frequencies, it could seem better to approximate their ratio as close as possible with a suitable pair of integers. However, beside possible computational problems, there are arguments on which a more effective choice can be based. Actually, the resonance should be of the lowest possible commensurability giving rise to a frequency ratio compatible with the dynamics of the actual system. The reason for this is that, the lower the order of the resonance, the richer the family of terms compatible with it that are available to construct the normal form. 

Moreover, another argument in favor of low order resonances comes from their role in the stability properties of periodic orbits. A typical situation is that in which a family of periodic orbits becomes unstable when a low order resonance occurs between its fundamental frequency and that of a normal perturbation: the simplest case is given by an axial orbit that, depending on the specific form of the potential, can be unstable through bifurcation of loop orbits (1:1 resonance), ``banana'' orbits (1:2 resonance), ``fish'' orbits (2:3 resonance),  etcetera. Therefore a detuned low-order resonant normal form can be quite accurate in describing the corresponding bifurcations. 

Finally, it must be emphasized that the structure of a resonant normal form is also affected by the symmetries of the original system. The normal form must preserve these symmetries and this in general also leads to a criterion for truncation. In the present instance of a double reflection symmetry, given a resonance ratio $m/n$, the normal form must contain at least terms of degree $2(m+n)$ (see, e.g. Tuwankotta and Verhulst, 2000). Therefore, the criterion we have adopted in this paper has been that of working with the lowest order truncated normal form incorporating the symmetries of a typical galactic potential: the 1:1 symmetric resonance which allows to truncate the normal form to degree 4 and the 1:2 symmetric resonance which requires to truncate the normal form to degree 6. A systematic investigation of the optimal order of truncation has recently been performed by Contopoulos et al. (2003) and Efthymiopoulos et al. (2004). Their results confirm the rapid decrease of the optimal order with the radius of the phase-space domain in which expansions are computed: we may conjecture that if we are interested in the global dynamics and accept a moderate level of accuracy, with this very conservative approach we can get reliable information up to the breakdown of the regular dynamics. 

\section{1:1 symmetric resonance and first order normalization}

A Lie transform normalization truncated to the second order gives the following expression of the first-order normal form (cfr. Belmonte et al. 2006)
$$\begin{array}{llll}
&& K_{2}^{(1:1)} = \frac12 \delta (P_1^2+ Q_1^2) + \frac{3}{32} \left[b_{40}({P_1}^2 + {Q_1}^2)^2 + b_{04}({P_2}^2 + {Q_2}^2)^2 \right] \\ \\ 
  &&           + \frac{b_{22}}{32}   \left( {P_2}^2 (3 {P_1}^2 + {Q_1}^2) + {Q_2}^2  ({P_1}^2 + 3                   {Q_1}^2) + 4 P_1 P_2 Q_1 Q_2 \right) ,\\ \\
&& K_{2}^{(m:n)} = \frac{n}{2} \delta (P_1^2+Q_1^2) + k_{1} \left[b_{40}({P_1}^2 + {Q_1}^2)^2 + b_{04}({P_2}^2 + {Q_2}^2)^2\right] \\ \\
&&  \quad \quad \quad + k_{2} \left[ b_{22}({P_1}^2 + {Q_1}^2) ({P_2}^2 + {Q_2}^2) \right],  \\ \nonumber \\ 
\end{array}
$$
where $k_{1}, k_{2}$ are rational numbers dependent on $m$ and $n$, Eq. (\ref{DET}) has been used, that in the present instance reads
\be\label{DET11}
\omega_1 = (1+ \varepsilon^2 \delta) \omega_2\ee 
and the canonical variables ${\bf P},{\bf Q}$ are as in Eq. (\ref{HK}). We see that in this case we have the same situation as in the first order averaging approach (Verhulst, 1996): the 1:1 resonance {\it or all other resonances}. 

To show more clearly how the symmetries influence the structure of the normal form, we have that it can be written as
\be\label{HDZM}
K = J_1 +  J_2 +  \varepsilon^2
[\delta J_{1} + \frac{3}{8} (b_{40} J_1^2 + b_{04} J_2^2 + \frac{2}{3} b_{22} J_1 J_2 ( 2 + \cos(2 \theta_1 - 2 \theta_2) ) )]\ee
where an overall rescaling by a factor $\omega_2$ has been performed and the action--angle variables are introduced according to
\ba\label{AAV}
Q_1 &=& \sqrt{2 J_1} \sin \theta_1,\\
P_1 &=& \sqrt{2 J_1} \cos \theta_1,\\
Q_2 &=& \sqrt{2 J_2} \sin \theta_2,\\
P_2 &=& \sqrt{2 J_2} \cos \theta_2.\ea
In fact, inverting these expressions and putting them into (\ref{HDZM}) we get
\be\label{HAVNF}
\omega_2 K = H_{0} + \omega_2 \varepsilon^2 K_{2}^{(1:1)}.\ee
The structure of (\ref{HDZM}) displays the effect of the symmetries on the resonant part: angles appear only through the combination $2 \theta_1 - 2 \theta_2$ and this shows why the symmetric 1:1 resonance can also be dubbed a ``2:2'' resonance. 
 
We can use (\ref{HDZM}) to identify the main periodic orbits. The procedure is the following (Sanders and Verhulst, 1985, sect.7.4). We perform the following canonical transformation to ``adapted resonance coordinates''
\ba
\psi &=& 2 (\theta_1 - \theta_2), \label{psi}\\
\chi &=& 2 (\theta_1 + \theta_2), \label{chi}\\
J_1 &=& ({\cal E} + {\cal R})/2, \label{j1}\\
J_2 &=& ({\cal E} - {\cal R})/2. \label{j2}\ea
In this way we get
\be\label{KER}
{\widetilde K} = 
\frac12 \delta ({\cal E} + {\cal R}) + A({\cal E}^{2} + {\cal R}^{2}) + B{\cal E}{\cal R} +C ({\cal E}^{2} - {\cal R}^{2})  (2 + \cos \psi) \ee
where the new action ${\cal E}$ is the additional integral of motion and has therefore been subtracted to get the effective Hamiltonian
\be
{\widetilde K} = \frac{K- {\cal E}}{\varepsilon^2}.\ee 
The coefficients 
\ba
A &=& \frac3{32} (b_{40}+b_{04}) ,\label{muA}\\
B &=& \frac3{16} (b_{40}-b_{04}) , \label{muB}\\
C &=& \frac{1}{32}b_{22},\label{muC}\ea
appearing in ${\widetilde K}$, give it the simplest form.

Considering the dynamics at a fixed value of ${\cal E}$, we have that ${\widetilde K}$ defines a one--degree of freedom $(\psi,{\cal R})$ system. We get the following equations of motion
\ba
{\dot \psi} &=& {\widetilde K}_{\cal R} = 
\frac12 \delta + B{\cal E} +2\left(A - C (2 + \cos \psi)\right){\cal R},\label{dpsi}\\
{\dot {\cal R}} &=& - {\widetilde K}_{\psi} = 
C \left({\cal E}^{2}  - {\cal R}^{2} \right) \sin \psi.\label{dr}
\ea
Let us determine the fixed points of this system: these in turn give the periodic orbits of the original system. The right hand of (\ref{dr}) vanishes either for ${\cal R}=\pm {\cal E}$ or for $\psi = 0, \pm \pi$. In the first case, the right hand of (\ref{dpsi}) vanishes when
\be\label{psiI}
\delta + 2 \left[B \pm 2 \left(A - C (2 + \cos \psi)\right) \right]{\cal E}=0\ee
and the two periodic orbits
\ba
{\cal R}&=&{\cal E}, \;\;\;\;  J_{2}=0, \quad ({\rm Type \; Ia}), \label{Ia}\\ 
{\cal R}&=&-{\cal E}, \; J_{1}=0, \quad ({\rm Type \; Ib}), \label{Ib}\ea
ensue. In the second case, the right hand of (\ref{dpsi}) vanishes either when
\be\label{psizero}
{\cal R} = \frac{\delta + 2B  {\cal E} }{4 (3C-A)}, \;\;(\psi=0),\ee
or when
\be\label{psipi}
{\cal R} = \frac{\delta + 2B {\cal E} }{4 (C-A)}, \;\;(\psi=\pi).\ee
The fixed point in (\ref{psizero}) determines the ``inclined'' orbit
\be
J_1 = \frac{\delta + 2 (B+2(3C-A)) {\cal E}}{8 (3C-A)}, \; ({\rm Type \; II}).\ee
Note that 
\be\label{range}
0 \le J_1 \le {\cal E}\ee
and this range determines the condition for existence of the orbit of Type II. The fixed point in (\ref{psipi}) determines the elliptic orbit
\be
J_2 = \frac{\delta + 2 (B+2(C-A)) {\cal E}}{8 (C-A)}, \; ({\rm Type \; III}).\ee
The range (\ref{range}) still determines the condition for existence of the orbit of Type III. 

Let us now consider the question of the stability of the periodic orbits. In particular, we are interested in what happens in the case of axial orbits of Type I: unfortunately, action--angle variables have singularities on these orbits and these affect also the adapted resonance coordinates. However, the remedy is quite straightforward: to use a mixed combination of action--angle variables on the normal mode and Cartesian variables for the other degree of freedom. The ensuing procedure is then first to determine the condition for the normal mode to be a critical curve of the Hamiltonian in these coordinates. Second, to assess its nature (Kummer, 1977; Contopoulos, 1978; Sanders and Verhulst, 1985, sect.7.4.4): the condition is found by considering the function
\be\label{LM1}
K^{(\mu)}=K+\mu H_{0},\ee
where $\mu$ has to be considered as a {\it Lagrange multiplier} to take into account that there is the constraint $ H_{0} = {\cal E}$. The Lagrange multiplier is found by imposing
\be\label{LM2}
{\rm d} \, K^{(\mu)} =0\ee
on the normal mode. Its nature is assessed by computing the matrix of the second derivatives of $K^{(\mu)}$: if the Hessian determinant of the second variation is positive definite the mode is elliptic stable; if it is negative definite the mode is hyperbolic unstable. 

In the case of the y-{\it axis} orbit of Eq. (\ref{Ib}), good coordinates are 
\ba\label{AAW1}
Q_1 &=& X,\\
P_1 &=& U,\\
Q_2 &=& \sqrt{2 J} \sin \theta,\\
P_2 &=& \sqrt{2 J} \cos \theta,\ea
so that the periodic orbit is given by 
\be\label{TIb}
X=U=0, \;\; J={\cal E}.\ee 
The terms in the normal form are then
\be
H_{0} = \frac12 (X^{2}+U^{2})+J\ee
and
\be\begin{array}{ll}
K_{2} = & \frac12 \delta (X^{2}+U^{2}) + \frac{3}{32} \left[b_{40}(X^{2}+U^{2})^2 + 4b_{04}J^2 \right] +\\  
      & \frac{1}{16} b_{22}  J \left[ 2 (X^{2}+U^{2}) + (X^{2}-U^{2}) \cos 2\theta + 2 X U \sin 2\theta \right].        \end{array}\ee
It is straightforward to check that, in this case, imposing Eq. (\ref{LM2}) on the periodic orbit defined by  
Eq. (\ref{TIb}), we get the equation
\be
\mu + 1 + \frac34 b_{04} {\cal E} = 0,\ee
which allows to find the required value of the Lagrange multiplier. With this result, the matrix of the second derivatives of $K^{(\mu)}$ on the normal bundle to the periodic orbit is
\be \frac18
\left( \begin{array}{cc} 8\delta + {\cal E} [b_{22} (2+\cos 2\theta) - 6 b_{04}] & 
                                   {\cal E} b_{22} \sin 2\theta \\ 
                                   {\cal E} b_{22} \sin 2\theta &
                                   8\delta + {\cal E} [b_{22} (2-\cos 2\theta) - 6 b_{04}] \end{array} \right).\ee
The equation ${\rm det}K^{(\mu)}({\cal E})=0$ gives
\be\label{DETER}
(36 b_{04}^{2}-24 b_{04} b_{22}+3 b_{22}^{2}){\cal E}^{2}-32(3 b_{04} - b_{22}) \delta {\cal E}+ 64 \delta^{2}=0\ee
with roots 
\be
{\cal E}_{1}= \frac{8 \delta}{6 b_{04} - b_{22}}, \;\;
{\cal E}_{2}= \frac{8 \delta}{3(2 b_{04} - b_{22})}.
\ee
Recalling the rescaling in (\ref{HDZM}), the physical energy is given by
\be E=\omega_2 {\cal E}.\ee 
If, as in the application cases that will be examined later, the first coefficient in Eq. (\ref{DETER}) is positive, the range of instability of the y-{\it axis} orbit is
\be\label{diseDZMY}
\frac{8 \omega_2 \delta }{6 b_{04} - b_{22}}
<E<
\frac{8 \omega_2 \delta}{3(2 b_{04} - b_{22})}.
\ee
Proceeding in the same way in the case of the x-{\it axis} orbit of Eq. (\ref{Ia}), the analogous expressions 
\be\label{diseDZMX}
\frac83 \frac{\omega_2 \delta }{ b_{22}- 2 b_{40}}
<E<
\frac{8 \omega_2 \delta }{ b_{22}- 6 b_{40}}
\ee
can be written for the other conditions. 

\section{1:2 symmetric resonance and second order normalization}

In the case of the 1:2 resonance in presence of reflection symmetries about both axes, we know that the normal form must be pushed at least to degree $2 \times (1+2) = 6$. We therefore have to perform a further step of normalization and include $K_4$ in the normal form. Its expression is quite involved (cfr. Belmonte et al. 2006), but we can exploit the change of variables to action--angle coordinates to see that the normal form has the structure
 \be\label{HDZM12}
K = J_1 +  2 J_2 + \varepsilon^2 (2 \delta J_1 + P_2(J_1 , J_2)) + \varepsilon^4 (P_3 (J_1 , J_2) + k J_1^2 J_2 \cos(4 \theta_1 - 2 \theta_2)),\ee
where Eq. (\ref{DET}) has been used, that in the present instance reads
\be\label{DET12}
\omega_1 = \left(\frac12 + \varepsilon^2 \delta \right) \omega_2, \ee
the polynomials $P_2$ and $P_3$ are homogeneous of degree $2$ and $3$ respectively
\begin{eqnarray*}
P_{2} &=& \frac38 b_{40} J_{1}^2 + \frac14 b_{22} J_{1} J_{2}  + \frac38 b_{04} J_{2}^2  \\
P_{3} &=& -\frac{1}{384} \bigg((102 b_{40}^2 - 160 
      b_{60} ) J_{1}^3 + (10 b_{22}^2 + 72 b_{22} 
        b_{40} - 96 b_{42}) J_{1}^2 J_{2} \\
        &&  + (36 b_{04} b_{22} + 10 b_{22}^2 - 96 b_{24}) J_{1} J_{2}^2 + (51 b_{04}^2 - 160 b_{06}) J_{2}^3 \bigg)
\end{eqnarray*}
and
\be
k = -\frac{1}{192} \left(3 b_{22}^2 - 3 b_{22} b_{40} - 8 b_{42}\right).\ee
The study of existence and stability of periodic orbits proceeds in the same way as in the previous section: we limit the presentation to the axial orbits. In the present case they are given by
\ba
 {\cal R}&=&{2\cal E}, \;\;\;\;  J_{2}=0, \quad ({\rm Type \; Ia}), \label{IIa}\\ 
 {\cal R}&=&-{\cal E}/2, \; J_{1}=0, \quad ({\rm Type \; Ib}). \label{IIb}\ea
The procedure to determine the condition of stability of axial orbits lead to analyze critical curves of the function (\ref{LM1}) where $K$ is given by (\ref{HDZM12}). Considering the x-{\it axis} (type Ia, Eq. (\ref{IIa})) orbit, good coordinates are 
\ba\label{AAW2}
Q_1 &=& \sqrt{2 J} \sin \theta,\\
P_1 &=& \sqrt{2 J} \cos \theta,\\
Q_2 &=& Y,\\
P_2 &=& V,\ea
so that the periodic orbit is given by 
\be\label{TIIb}
Y=V=0, \;\; J={\cal E},\ee 
 and
\be\label{LM12}
H_{0} = J + Y^{2} + V^{2}.\ee
The condition Eq. (\ref{LM2}) allows to find the Lagrange multiplier
\be
\mu = -1 - 2 \delta - \frac{\cal E}{4} \left[ 3 b_{40} \left(1 - \frac{17}{16} b_{40} {\cal E} \right) + 5 b_{60} {\cal E} \right].\ee
The equation ${\rm det}K^{(\mu)}({\cal E})=0$, obtained by computing the matrix of the second derivatives of $K^{(\mu)}$ on the normal bundle to the periodic orbit (\ref{TIIb}), is 
\be\label{dise12X}
(A_{2} {\cal E}^{2} + A_{1} {\cal E} + 768 \delta )(A_{2}' {\cal E}^{2} + A_{1} {\cal E} + 768 \delta )=0,\ee
where
\ba
A_{1}&=&48 (6 b_{40} - b_{22}) ,\\
A_{2}&=&2 b_{22}^{2} + 33 b_{22} b_{40} - 306 b_{40}^2 - 
              56 b_{42} + 480 b_{60},\\
 A_{2}'&=& 2 b_{22}^{2} + 39 b_{22} b_{40} - 306 b_{40}^2 - 
              40 b_{42} + 480 b_{60}.\ea
      We see that the inequalities to be satisfied in this case are quadratic in the energy, contrary to the linear case of (\ref{diseDZMY}) and (\ref{diseDZMX}). Below we discuss the specific example of the x-{\it axis} orbit in the logarithmic potential.
      

\section{Approximate integral in the original variables}

The general procedure illustrated so far can be applied to study the dynamics associated to every reduced normal form. Specifically, it can be extended to forms truncated to an arbitrary high order and to all normal modes and periodic orbits associated to the given resonance. However, if the 1:1 and  1:2 cases treated here are rather simple, computations become quite cumbersome for higher orders. It can be useful to exploit an alternative, less general, but easier procedure. 
Recalling the generic expression of the second invariant (\ref{eqn:In}), if we truncate at first order, we have
\be\label{inva2}
I_{2} = H_{0}+\varepsilon^2(V_{2}-K_{2}).\ee
This is the best approximate integral of motion of a symmetric perturbed 1:1 oscillator to order $\varepsilon^4$ in the perturbation parameter: in fact, due to the symmetries of the problem, odd-degree terms are absent both in the normal form and in the approximate integral. To assess the stability of, e.g., the y-{\it axis} orbit, we may proceed in this way: to use $I_2$ and the conserved energy
\be
E =  \frac{1}{2}(p_x^2+p_y^2) + V_{0}(x^2,y^2)+\varepsilon^2 V_2 (x^2,y^2)\ee
to construct an $x-p_x$ Poincar\'e section by means of the intersection of the function $I_2 (x,y,p_x;E)$ with the $y=0$ hyperplane and the level curves of the function
\be\label{FF}
F = I_2 (x,0,p_x;E);\ee
to study the nature of the origin as a critical point determining if it is either an extremum (elliptic fixed point = stable periodic orbit) or a saddle (hyperbolic  fixed point = unstable periodic orbit) by using the Hessian determinant
\be F_{p_x p_x} F_{x x} - F_{x p_x}^{2}.\ee 
Clearly, for resonant periodic orbits even the location of the critical points can be already quite difficult and this limits the generality of the approach. However, in the case of axial orbits, the approach is straightforward: in the case of the y-{\it axis} orbit, we have that the second derivatives in the origin are \ba
F_{p_x p_x} = 4 \omega_2 \delta - 3 E 
\left((1+\delta)b_{04}-\frac12 b_{22}\right) ,\\
F_{x x} = 4 \omega_2 \delta - E 
\left(3(1+\delta)b_{04}-\frac12 b_{22}\right) .\ea
The mixed second derivative $F_{x p_x}$ is identically zero in the origin. The range of instability (namely, the range where the two second derivatives have different sign) is
\be\label{diseINV}
\frac{8 \omega_2 \delta }{ 6 b_{04} (1+\delta) - b_{22}}
<E<
\frac{8 \omega_2 \delta}{3 (2 b_{04}  (1+\delta)- b_{22})}.
\ee
Comparing with (\ref{diseDZMY}) we have an order $\delta^{2}$ discrepancy whose origin is connected with the construction of the function (\ref{FF}) where the term with the detuning is not considered as a perturbation. Analogous inequalities can be written for the x-{\it axis} orbit by constructing a $y-p_y$ Poincar\'e section and studying the level curves of the function
\be\label{FFY}
F = I_2 (0,y,p_y;E)\ee 
obtained by means of the intersection of the function $I_2 (x,y,p_y;E)$ with the $x=0$ hyperplane.

      On the other side, we can develop this parallel approach of determining the nature of fixed points on the surface of section for higher order approximate integrals too. Using  (\ref{eqn:In}), if we truncate at second order, we have
\be\label{inva4}
I_{4} = H_{0}+\varepsilon^2(V_{2}-K_{2})+\varepsilon^4(V_{4}-\{g_{2},K_{2}\}-K_{4})\ee
where $g_{2}$ is the first order generating function determined in the first step of the normalization. This is the best approximate integral of motion of a symmetric perturbed 1:2 oscillator to order $\varepsilon^6$ in the perturbation parameter. Let us consider the x-{\it axis} orbit: the construction of the $y-p_y$ Poincar\'e section and the study of the critical points of the function
\be\label{FFFY}
F = I_4 (0,y,p_y;E),\ee 
obtained by means of the intersection of the function $I_4 (x,y,p_y;E)$ with the $x=0$ hyperplane,  proceeds as above. Concerning the nature of the fixed point in the origin (linked to the stability of the x-{\it axis} orbit), we get quadratic inequalities in the energy analogous to those arising from (\ref{dise12X}): the discrepancy between the two methods  is still of order $\delta^{2}$ and is discussed in the subsequent section about a specific example.


\section{Comparison of analytical and numerical results}

We have applied the theory discussed in the previous sections to the logarithmic and the Schwarzschild potentials. We can compare these analytical predictions concerning the stability energy ranges of axial orbits with other results, mainly numerical, from the literature. In particular, we have chosen the work by Miralda-Escud\'e and Schwarzschild (1989, MES in what follows) who made an accurate numerical exploration of the phase-space structure of the logarithmic potential: they give the existence parameter ranges of the main families of periodic orbits and the bifurcation ensuing from instability thresholds in two models (corresponding to ellipticity values $q=0.7$ and $q=0.9$) determined by solving the perturbation equations with the Floquet method. The authors use the core radius $R_{c}$ to parametrize the sequence of periodic orbits because they were interested in comparing the results with the case of the {\it singular} scale--free case $R_{c}=0$: in fact they fix the energy $E=0$ in all their computations and vary $R_{c}$ and $q$. For our purposes, it is more natural to use the energy as a parameter and therefore we have made the conversion 
\be
E = - \log R_{c}\ee
to compare the results. Another approach to the study of the stability of axial orbits in 
the logarithmic potential has been followed by Scuflaire (1995, SC in what follows) who solved the Hill--like equation of the perturbation by means of a Poincar\'e--Lindstedt series expansion up to order 20. SC uses $a$, the amplitude of the axial orbit, as a parameter: the conversion to the energy is given by
\be
E = \frac12 \log (1+a^{2}).\ee
In tables I. and II. we display the results in the case of the logarithmic potential: in the first column are the two values of the parameter $q$ for which reliable numerical estimates are available; in the second are the analytical predictions according to the procedures of Section 3 for table I and 4 for table II; in the third are the analytical predictions according to the procedures of Section 5; in the fourth the numerical prediction of MES; in the last the semi--analytical prediction of SC. 

The values shown in table I. represent the energy thresholds to instability of the short y-{\it axis} orbit: this orbit becomes unstable when its frequency falls to that of a normal perturbation and as soon as this happens a loop orbit bifurcates. The values of the second and third columns are therefore computed using the results of the 1:1 resonance treatment of Sections 3 and 5, namely the first inequality in (\ref{diseDZMY}) and the first in (\ref{diseINV}). The computations are performed by using the coefficients given in (\ref{alog2}) recalling definition (\ref{abc}). Clearly the accuracy of both predictions is better for $q=0.9$ corresponding to a smaller detuning of the 1:1 resonance. Overall, we see that the prediction using the invariant curves performs significantly better than that obtained directly working with the normal form. 

\begin{table}
    \begin{tabular}{||c||c|c||c|c||}
      \hline
 $q$ & ${\rm Normal - Form}$ &      ${\rm Invariant}$ &  ${\rm MES}$ & ${\rm SC}$ \\
									   \hline
 $0.7$ & $ 0.52 $  &   $ 0.86 $ &  $ 0.72 $ & $ 0.80 $ \\
 $0.9$ & $ 0.19 $  &   $ 0.22 $ &  $ 0.21 $ & $ 0.25 $ \\
 \hline
\end{tabular}
  \caption{Energy threshold of instability for the short y-{\it axis} orbit in the logarithmic potential}\label{tab:y_{axis}}
\end{table}

 \begin{table}
    \begin{tabular}{||c||c|c||c|c||}
      \hline
 $q$ & ${\rm Normal - Form}$ &      ${\rm Invariant}$ &  ${\rm MES}$ & ${\rm SC}$ \\
									   \hline
 $0.7$ & $ 1.32 \div 4.82 $  &   $ 1.71 \div 4.86 $ &  $ 1.52 \div 4.29 $ & $ 1.42 \div - $ \\
 $0.9$ & $ 0.85 $  &   $ 0.80 $ &  $ 3.64 $ & $ - $ \\
 \hline
\end{tabular}
  \caption{Energy intervals of instability for the long x-{\it axis} orbit in the logarithmic potential}\label{tab:x_{axis}}
\end{table}

\begin{table}
    \begin{tabular}{||c|c||c|c||}
      \hline
 ${\rm Normal - Form}$ & ${\rm Invariant}$ &  ${\rm DZM \; averaging}$ & ${\rm DZM \;numerical}$ \\
									   \hline
 $0.19$ & $ 0.28 $  &   $ 0.21 $ &  $ 0.27 $ \\
 \hline
\end{tabular}
  \caption{Energy threshold of instability for the short y-{\it axis} orbit in the Schwarzschild potential}\label{tab:schy_{axis}}
\end{table}

The long x-{\it axis} orbit can suffer instability, but not through a 1:1 resonance. This orbit becomes unstable when its frequency falls to $1/2$ of that of a normal perturbation and, as soon as this happens, a banana orbit bifurcates. Therefore, we turn the attention to table II. where the values of the second and third columns are computed using the results of the 1:2 resonance treatment of Sections 4 and 5: namely, the condition in order to the quartic in (\ref{dise12X}) is negative, using the coefficients computed from (\ref{alog2}) and (\ref{alog3}) and the nature of the fixed point in the origin as determined through function (\ref{FFFY}). An important prediction of the numerical approaches is that, if the ellipticity is less than a given value, there is an instability {\it interval} in energy: actually in the figure of SC it is not possible to determine the upper extremum of it, but this interval is very well determined in the computations of MES. We see that the analytical estimates of the instability interval are not very far from them, with an acceptable prediction of the upper boundary of the interval too. Our treatment also provides a critical value of $q_{c}=0.78$ of the ellipticity, beyond which the interval disappears. It would be interesting to numerically test this prediction. For higher values of $q$ the analytical estimates of the threshold of instability are not so good and the reason is that the detuning of the 1:2 resonance becomes excessively large.

 Finally, in table III. we report the comparisons concerning the short y-{\it axis} orbit in the Schwarzschild potential, with the coefficients listed in (\ref{SA1}) and (\ref{SA2}): in the third column we report the prediction given by de Zeeuw and Merritt (1983, DZM) using the first order averaging of the equations of motion and in the fourth the corresponding prediction of the same authors using a numerical integration. We see that, as usual, the analytical prediction based on the invariant is quite accurate. 
 
 We conclude observing that applications like those presented here confirm that this very conservative strategy provides sufficient qualitative and quantitative agreement with other more accurate but less general approaches. An important issue to clarify is if, in order to improve the degree of accuracy, it is enough to simply truncate the normal form to higher orders or if it is rather necessary to incorporate in a more careful way the detuning in higher order terms of the perturbation.

\acknowledgements  Our thanks to Giuseppe Gaeta and Ferdinand Verhulst for very fruitful discussions.

\end{article}

\begin{thebibliography}{10}
 
\bibitem{noi}
Belmonte C. Boccaletti D. and Pucacco G. (2006) ``Approximate First Integrals for a Model of Galactic Potential with the Method of Lie Transform Normalization'', submitted to the Proceedings of the Saarifest 2005, E. Perez-Chavela and J. Xia editors.

\bibitem{bir}
Birkhoff G. D. (1927) {\it Dynamical Systems}, Amer. Math. Soc. 
Coll. Publ., {\bf 9}, New
York, USA

     \bibitem{BT}
   Binney J. and Tremaine S. (1987) {\it Galactic Dynamics}, Princeton University Press.
   
   \bibitem{DB}
   Boccaletti D. and Pucacco G. (1999) {\it Theory of Orbits: Vol. 2}, Springer-Verlag, Berlin.
  
  \bibitem{cont}
  Contopoulos G. (1978) ``Higher order resonances in dynamical systems''.  {\it Cel. Mech.}, {\bf 18}, 195--204.
  
\bibitem{conto}
Contopoulos G. (2002) {\it Order and Chaos in Dynamical Astronomy}, Springer-Verlag, Berlin.

  \bibitem{CEG}
     Contopoulos G. Efthymiopoulos C. and Giorgilli A. (2003) ``Nonconvergence of formal integrals of motion''. {\it  J. Phys. A: Math. Gen.}, {\bf 36}, 8639--8660.
     
\bibitem{zm}
   de Zeeuw T. and Merritt D. (1983) ``Stellar orbits in a triaxial galaxy. I. Orbits
in the plane of rotation''. {\it Astrophys. J.}, {\bf 267}, 571--595.

 \bibitem{DF}
   Dragt A. and Finn J. M. (1976) ``Lie series and invariant functions for analytic
symplectic maps''. {\it J. Mat. Phys.}, {\bf 17}, 2215--2227.
    
      \bibitem{ECG}
     Efthymiopoulos C. Giorgilli A. and Contopoulos G. (2004) ``Nonconvergence of formal integrals: II. Improved estimates for the optimal order of truncation''. {\it  J. Phys. A: Math. Gen.}, {\bf 37}, 10831--10858.
     
   \bibitem{FINN}
   Finn J. M. (1984) ``Lie series: a perspective. Local and Global Methods of Nonlinear Dynamics''. {\it Lecture Notes in Physics}, {\bf 252}, 63--86.

\bibitem{fm}
Fridman T. and Merritt D. (1997) ``Periodic orbits in triaxial galaxies with weak cusps''. {\it Astron. J.}, {\bf 114}, 1479--1487.

\bibitem{gu}
Gustavson F. (1966) ``On constructing formal integrals of a Hamiltonian system
near an equilibrium point''. {\it Astron. J.}, {\bf 71}, 670--686.

   \bibitem{KO}
   Koseleff P. V. (1994) ``Comparison between Deprit and Dragt-Finn perturbation methods''.  {\it Cel. Mech. \& Dynam. Astron.}, {\bf 58}, 17--36.
  
   \bibitem{KU}
   Kummer M. (1977) ``On resonant Hamiltonians with two degrees of freedom near an equilibrium point''. {\it Lecture Notes in Physics}, {\bf 93}, 57--75.
   
  \bibitem{mes}
  Miralda-Escud\'e J. and Schwarzschild M. (1989) ``On the orbit structure of the
logarithmic potential''. {\it Astrophys. J.}, {\bf 339}, 752--762.
  
  \bibitem{moser}
  Moser J. (1968) ``Lectures on Hamiltonian systems''. {\it Mem. Am. Math. Soc.}, {\bf 81}, 1--60.
  
  \bibitem{SV}
Sanders J. A. and Verhulst F. (1985) {\it Averaging Methods in Nonlinear Dynamical Systems}, Springer-Verlag, New York.

  \bibitem{sch}
  Schwarzschild M. (1979) ``A numerical model for a triaxial stellar system in dynamical equilibrium''. {\it Astrophys. J.}, {\bf 232}, 236--247.

\bibitem{scu}
Scuflaire R. (1995) ``Stability of axial orbits in analytic galactic potentials''.  {\it Cel. Mech. \& Dynam. Astron.}, {\bf 61}, 261--285.

   \bibitem{tv}
   Tuwankotta J. M. and Verhulst F. (2000) ``Symmetry and Resonance in Hamiltonian Systems''. {\it SIAM J. Appl. Math.}, {\bf 61}, 1369--1385.

\bibitem{ver}
   Verhulst F. (1996) {\it Nonlinear Differential Equations and Dynamical Systems}, Springer-Verlag, Berlin.
   
\end{thebibliography}
\end{document}